\documentclass[aps,prl,preprint,amsfonts]{revtex4-1}

\usepackage{amsmath}
\usepackage{graphics}
\usepackage[english]{babel}
\usepackage{multirow}	  
\usepackage{rotating}     

\begin{document}

\title{Analytical Determination of Fractal Structure in Stochastic Time Series}
\date{\today}
\author{Ferm\'{\i}n \surname{Moscoso del Prado Mart\'{\i}n}}
\email[Contact: ]{fermin.moscoso-del-prado@univ-provence.fr}
\affiliation{Laboratoire de Psychologie Cognitive ({UMR}--6146)\\
{CNRS} \& Aix--Marseille Universit\'e I, Marseille, France}

\begin{abstract}
Current methods for determining whether a time series exhibits fractal structure (FS) rely on subjective assessments on estimators of the Hurst exponent (H). Here, I introduce the Bayesian Assessment of Scaling, an analytical framework for drawing objective and accurate inferences on the FS of time series. The technique exploits the scaling property of the diffusion associated to a time series. The resulting criterion is simple to compute and represents an accurate characterization of the evidence supporting different hypotheses on the scaling regime of a time series. Additionally, a closed-form Maximum Likelihood estimator of H is derived from the criterion, and this estimator outperforms the best available estimators.
\end{abstract}
\keywords{$1/f^\alpha Noise; Bayesian Assessment of Scaling; Diffusion; Fractal; Hurst Exponent; Scaling;}

\maketitle

\newcommand{\ud}{\,\mathrm{d}}

In this study I introduce a method for deciding whether a time series exhibits fractal scaling  ({FS}; \cite{Hurst:1951,Mandelbrot:VanNess:1968,Mandelbrot:Wallis:1969}) in a simple, objective, and accurate way. Determining the presence or absence of FS in a time series is an important question for making inferences about the nature of underlying system that generated it. Important conclusions concerning the nature of systems in a wide range of domains rest on assessing whether -- and, if so, the extent to which -- observed time series exhibit FS. Usually, this is addressed by estimating whether the value of the Hurst exponent (H; \cite{Mandelbrot:Wallis:1969}) of the series is equal to or different from 1/2. Often, some additional verification is performed by looking for the consistency of multiple estimators, but the criteria for establishing FS retains a fair deal of subjectivity (see, {\em e.g.}, \cite{Eke:etal:2002} for a heuristic methodology for combining estimators). However, estimating the value of a parameter does not constitute a valid assessment of FS. Rather, in order to argue that a series is fractal (or, more generally, its H exponent ranges within some hypothesized interval) an explicit contrast between different hypotheses about H is required (see \cite{Jaynes:2003}, for a discussion distinguishing parameter estimation from hypothesis contrast). The few approaches that consider some hypothesis testing rely on approximative techniques with multiple adjustable parameters and strong assumptions on the underlying model \cite{Davies:Harte:1987,Wagenmakers:etal:2004,Wagenmakers:etal:2005,Thornton:Gilden:2005}.

The elements in a time series can be thought of as the fluctuations of a particle following a diffusion process \cite{Peng:etal:1994}. From this perspective, for any process $X(\tau)$ generating values with a stationary distribution $\mathrm{f}_X$, one can consider the diffusion process $Y(\tau)$ associated to $X(\tau)$, which is defined by $\ud Y(\tau) / \ud \tau = X(\tau)$. Since what is normally available is not the continuous evolution of $X$, but rather a discretely sampled time series $\boldsymbol{x} = \{x(1), \ldots,x(N)\}$, one can instead define the corresponding random walk $\boldsymbol{y}$ on discrete time $k$
\begin{equation}
y(k) = \sum_{i=1}^{k} x(i),\qquad k = 1, \ldots, N.
\end{equation}
If the generating process $X$ is stationary, the \emph{scaling property} \cite{Allegrini:etal:1996} indicates that, for large times, the probability distribution of the walk approaches
\begin{equation}
p(y|k,\delta) = \frac{1}{k^\delta} \mathrm{f}_L\left(\frac{y}{k^\delta}\right),
\label{eq:scaling1}
\end{equation}
where $\delta$ is the \emph{scaling exponent} associated with the time series and $\mathrm{f}_L$ is a limiting probability density function. The scaling exponent is closely related to the H exponent \cite{Allegrini:etal:1996,Scafetta:Grigolini:2002}. In the cases when $\mathrm{f}_X$ corresponds to the normal distribution -- fractional Gaussian noise (fGn; \cite{Mandelbrot:VanNess:1968,Mandelbrot:Wallis:1969}) -- $\delta$ is equivalent to H.

Let $\boldsymbol{x} = \{x(1), \ldots ,x(N)\}$ be a stationary time series of $N$ samples, and $\boldsymbol{y}_k $ the $k$-th order running sum of $\boldsymbol{x}$, $y_k(i) = \sum_{j=i}^{i+k-1} x(j)$. The running sums can be taken as samples at time $k$ of an ensemble of $N-1$ trajectories from the diffusion process (see Refs.~\cite{Scafetta:Grigolini:2002, Ignaccolo:etal:2004a} and references therein). For large summing orders ({\em i.e.}, diffusion times), the scaling condition should apply. The simplest case is when two {\em a priori} candidate values for the scaling exponent, $\delta_1$ and $\delta_2$ are available. One's task is to determine which of the two scaling regimes better describes the complexity of the observed series $\boldsymbol{x}$. This  amounts to determining the Bayes factor $\lambda_{1,2} = \ln [\mathrm{p}(H_1 | \boldsymbol{x}) / \mathrm{p}(H_2 | \boldsymbol{x})]$ between two simple hypotheses $H_1 \equiv \delta = \delta_1$, and $H_2 \equiv \delta = \delta_2$. If $\lambda_{1,2} > 0$, then the data support hypothesis $H_1$ more strongly than hypothesis $H_2$, and the reverse holds if $\lambda_{1,2} < 0$. By Bayes' Theorem, $\lambda_{1,2}$ can be decomposed into the sum of the \emph{a priori} information $\alpha_{1,2}$ (obtained by theoretical predictions or previous observations of the same process) and the evidence provided by the data $\eta_{1,2}(\boldsymbol{x})$
\begin{equation}
\lambda_{1,2}(\boldsymbol{x}) = \ln \frac{\mathrm{p}(H_1)}{\mathrm{p}(H_2)} + \ln \frac{\mathrm{p}(\boldsymbol{x}|H_1)}{\mathrm{p}(\boldsymbol{x}|H_2)} = \alpha_{1,2} + \eta_{1,2}(\boldsymbol{x}).
\label{eq:Bayes1-}
\end{equation}
Exploiting the scaling property in Eq.~\ref{eq:scaling1}, the likelihood above can be expressed in terms of the $k$-th running sum $\boldsymbol{y}_k$. When doing this, one should consider that the overlap between running sum windows introduces much redundancy in the combined evidence. This imposes rescaling it from its length of $N-k+1$ elements, to its maximum effective length, $N/k$. The combined evidence is therefore
\begin{equation}
\eta_{1,2}(\boldsymbol{x}|k) = \frac{N}{(N-k+1)k} \sum_{i=1}^{N-k+1}\ln\frac{\mathrm{p}(y_k(i)|\delta=\delta_1,k)}{\mathrm{p}(y_k(i)|\delta=\delta_2,k)}.
\label{eq:sumw}
\end{equation}
Combining the scaling condition of Eq.~\ref{eq:scaling1} with the evidence of Eq.~\ref{eq:sumw} results in
\begin{align}
\eta_{1,2}(\boldsymbol{x}|k) & = \frac{N}{(N-k+1)k}\sum_{i=1}^{N-k+1} \ln \frac{\mathrm{f}_L\left[y_k(i)/k^{\delta_1}\right]}{\mathrm{f}_L\left[y_k(i)/k^{\delta_2}\right]} + \ldots \nonumber\\
& \quad + \frac{N}{k}(\delta_2 - \delta_1) \ln k.
\label{eq:sumscaling}
\end{align}
For a time series of length $N$, there are $N-1$ summing orders than can be investigated. The combined evidence across all orders corresponds to the weighted mean of the evidences across summing orders -- note that, as all the running sums were computed from a single realization of the series, a plain sum would exaggerate the evidence --

\begin{equation}
\eta_{1,2}(\boldsymbol{x}) = \frac{2}{N(N-1)}\sum_{k=2}^{N} (N-k+1)\eta_{1,2}(\boldsymbol{x}|k).
\label{eq:evidence2general}
\end{equation}

In a more general case when a candidate value for the scaling exponent $\delta$ is not available, but rather a range of possible values is provided, our task consists in comparing the simple hypothesis $H_0 \equiv \delta = \delta_0$ and the composite hypothesis $H_1 \equiv \delta \in [\delta_1,\delta_2]$. As before, the evidence for each hypothesis is expressed in terms of its $k$-order running sum. In the likelihood for the composite Hypothesis $H_1$, $\delta$ is a now a `nuisance' parameter, and the estimation of the likelihood requires integrating it away,
\begin{subequations}
\begin{align}
\mathrm{p}\left(\boldsymbol{y} | \delta \in [\delta_1, \delta_2],k\right)  & = \int_{\delta_1}^{\delta_2} \mathrm{p}\left(\boldsymbol{y} | \delta,k\right) \mathrm{p}\left(\delta | k\right) \ud \delta \label{eq:integralp1}\\
& = \frac{1}{\delta_2 - \delta_1} \int_{\delta_1}^{\delta_2} \mathrm{p}\left(\boldsymbol{y} | \delta ,k\right) \ud \delta \label{eq:integralp2}
\end{align}
\end{subequations}
Assuming nothing about the value of $\delta$ apart from it falling in the interval $[\delta_1,\delta_2]$, requires the use of an uninformative uniform prior for it, $\mathrm{p}\left(\delta = h | k\right) = 1/(\delta_2 - \delta_1)$, changing Eq.~\ref{eq:integralp1} into Eq.~\ref{eq:integralp2}. Combining the scaling condition with the log-likelihoods of the hypotheses, and integrating Eq.~\ref{eq:integralp2}, the expression for the evidence provided by summing order $k>1$ becomes
\begin{widetext}\begin{equation}
\eta(\boldsymbol{x}|k) = \frac{N}{k}\ln \frac{k^{\delta_0}}{\delta_2 - \delta_1} + \frac{N}{(N-k+1)k}\sum_{i=1}^{N-k+1} \ln \frac{\mathrm{F}_L\left[y(k,i)/k^{\delta_1}\right] -\mathrm{F}_L\left[y(k,i)/k^{\delta_2}\right]}{y(k,i)\mathrm{f}_L\left[y(k,i)/k^{\delta_0} \right] \ln k},
\label{eq:ll12}
\end{equation}\end{widetext}
where $\mathrm{F}_L(x) = \int_{-\infty}^x \mathrm{f}_L(z) \ud z$ is the cumulative probability function of $\mathrm{f}_L$ \footnote{The case comparing two composite hypotheses can be derived trivially from Eq.~\ref{eq:ll12}.}.

\begin{figure}
\includegraphics[scale=.45]{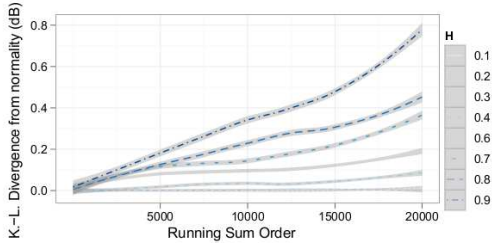}
\caption{\label{fig:kld} (color online) Deviation from normality, numerically estimated KLD between the Gaussian distribution ($\mathrm{f}_X$) and the distribution of the running sums of increasing order. For each value of $H$, the points were estimated using simulated {fGn} of $5\cdot 10^5$ elements.}
\end{figure}

Calculating the evidences in Eqs.~\ref{eq:sumscaling}, and \ref{eq:ll12} requires an analytical expression for the limiting distribution $\mathrm{f}_L$. Strictly speaking, the scaling condition applies exactly only in the infinite time limit. The process of convergence to this limit is one by which the original distribution of the data, $\mathrm{f}_X$, is gradually being transformed into the limiting distribution $\mathrm{f}_L$. Fig.~\ref{fig:kld} illustrates this point. It plots the departure from normality of the diffusion associated to an fGn as time increases. As expected, one finds a progressive increase in the Kullback-Leibler Divergence (KLD; \cite{Kullback:Leibler:1951}) between $\mathrm{f}_X$ and the observed distribution. Even for this simple fGn case, full convergence to  $\mathrm{f}_L$ is not attained at very large summing orders, but the divergence from normality is rather small.

By the scaling property, if $\delta$ is the hypothesized scaling exponent, the expected log-likelihood of the sum $\boldsymbol{y}_k$ is \footnote{Eq.~\ref{eq:step1} is transformed into Eq.~\ref{eq:step2}  by a simple change of variable $z = y/k^{\delta},\, \ud y = k^{\delta} \ud z$.}
\begin{subequations}
\begin{align}
\left<\ell\left[\boldsymbol{y}|k,N,\delta]\right]\right>_{\mathrm{f}_L} & =   \frac{N}{k} \int_{-\infty}^{\infty} \frac{1}{k^{\delta}} \mathrm{f}_L\left(\frac{y}{k^{\delta}}\right) \ln \mathrm{f}_L\left(\frac{y}{k^{\delta}}\right)\ud y - \ldots \nonumber\\
& \qquad -\frac{N}{k} \delta \ln k \label{eq:step1} \\
 & = -\frac{N}{k} \delta \ln k + \frac{N}{k} \int_{-\infty}^{\infty} \mathrm{f}_L(z) \ln \mathrm{f}_L(z) \ud z \label{eq:step2}.
 \end{align}
\end{subequations}
However, using $\mathrm{f}_X$ in place of $\mathrm{f}_L$ to estimate the likelihood would have resulted in an estimate
\begin{equation}
\left<\ell\left[\boldsymbol{y}|k,N,\delta]\right]\right>_{\mathrm{f}_X} = -\frac{N}{k} \delta \ln k + \frac{N}{k} \int_{-\infty}^{\infty} \mathrm{f}_L(z) \ln \mathrm{f}_X(z) \ud z,
\end{equation}
so that the estimation error would have been
\begin{align}
\Delta(k) & = \left<\ell\left[\boldsymbol{y}|k,N,\delta\right]\right>_{\mathrm{f}_L} - \left<\ell\left[\boldsymbol{y}|k,N,\delta]\right]\right>_{\mathrm{f}_X}  \nonumber \\
 & = \frac{N}{k}\int_{-\infty}^{\infty} \mathrm{f}_L(z) \ln \frac{\mathrm{f}_L(z)}{\mathrm{f}_X(z)} \ud z = \frac{N}{k} \mathrm{K}\left[\mathrm{f}_L \| \mathrm{f}_X\right],
\label{eq:klerror}
\end{align}
where $\mathrm{K}[\cdot\|\cdot]$ refers to the {KLD} between two distributions. Eq.~\ref{eq:klerror} reflects the upper bound of the error in estimating the log-likelihood. In general, for two summing orders $k_1 < k_2$, it should hold that $0 \leq k_1\Delta(k_1)/N \leq k_2\Delta(k_2)/N \leq \mathrm{K}\left[\mathrm{f}_L \| \mathrm{f}_X\right]$. It is therefore clear that using $\mathrm{f}_X$ instead of $\mathrm{f}_L$ to estimate the log-likelihood of the running sums should result in an underestimation of the likelihood, with the amount of underestimation bound by the KLD between the two distributions (the KLD between two distributions is always non-negative; \cite{Kullback:Leibler:1951}). In the {fGn} case, $\mathrm{f}_X$ is the Gaussian distribution and the approximation is exact for $\delta = 1/2$, and otherwise the underestimation increases with $\delta$. Therefore, comparing the approximated log-likelihoods for $\delta>1/2$ against the hypothesis that $\delta = 1/2$ will be slightly biased in favor of the latter null hypothesis (which is perhaps desirable).

In the case when one has sufficient evidence for the presence of FS, it is often necessary to determine what the scaling exponent is. A simple way to obtain such an estimate, is to find the value $\hat{\delta}$ that maximizes the posterior probability of the exponent given the observed data. If one assumes a uniform uninformative prior on $\delta$, this is equivalent to maximizing the log likelihood of the data given an exponent. Using the scaling condition provides an expression of the log-likelihood for each summing order ($\ell[\boldsymbol{x}|k,\delta]$) as a function of $\delta$. The maxima are found where the first derivative $\partial \ell / \partial \delta$ is zero with a negative second derivative $\partial^2 \ell / \partial \delta^2$ . If $\hat{\delta}(k)$ is a maximum of the likelihood then, assuming that that the log-likelihood is peaked around the single maximum $\hat{\delta}(k)$ with a shape that is roughly Gaussian, the variance of the estimator $\varepsilon(k)^2$ is given by the minus reciprocal of the second derivative evaluated at $\hat{\delta}(k)$. When $\mathrm{f}_X$ is a zero-mean Gaussian distribution of variance $\sigma^2$, one finds that the log-likelihood (again approximating $\mathrm{f}_L$ by $\mathrm{f}_X$) has a single maximum for all values of $k>1$ with
\begin{equation}
\hat{\delta}(k) = \frac{\ln\left( \sum_{i=1}^{N-k+1} y(k,i)^2\right) - \ln\left(\sigma^2(N - k + 1) \right)}{2\ln k}, \label{eq:MLest}
\end{equation}
and estimated error
\begin{equation}
\varepsilon(k)^2 =  \frac{(N-k+1)\sigma^2 k^{2\hat{\delta}(k)+1}}{2N\ln^2(k) \sum_{i=1}^{N-k+1} y(k,i)^2}. \label{eq:MLerror} 
\end{equation}
This produces $N-1$ estimators, one for each summing order. From these, the $\hat{\delta}(k)$ with the smallest error $\varepsilon(k)$ is chosen as the estimator.

\begin{figure}
\includegraphics[scale=.3]{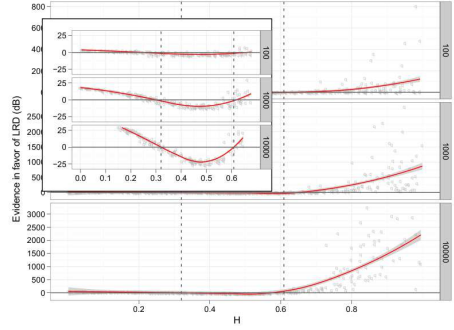}
\caption{\label{fig:evidencesFocus} (color online) Performance of the BAS criterion in detecting FS for 1,000 artificially generated fGn series of length 100 (top), 1,000 (middle), and 10,000 (bottom). The inset focuses on the lower values of $H$.}
\end{figure}

In order to explore the power and accuracy of the {BAS} criterion and the associated H estimator, I generated 1,000 artificial fGn series of lengths randomly chosen between 100, 1,000, and 10,000 elements, and real values of $\mathrm{H}=\delta$ uniformly sampled from the $(0,1)$ interval. Fig.~\ref{fig:evidencesFocus} plots the results of using the BAS criterion to compare the theory that $\delta \neq 1/2$ (or more precisely, $\delta \in [0,1]$) with the null hypothesis of $\delta = 1/2$. The figure shows that, when the real value of $\delta$ was either lower than $.32$ or higher than $.61$ (these limits are plotted by the vertical lines), the BAS clearly detects the FS. However, for values of $\delta$ within the interval $[.32,.61]$, the BAS supports the null hypothesis more strongly. This means that the theory of $\delta = 1/2$ presents a better description of the data than stating that nothing is known about $\delta$. Notice that this does not imply that $\delta = 1/2$ is the `correct' scaling, but rather, that it is the best among the theories compared. Choosing a better specified theory to test against, would enable contrasting hypotheses with higher resolution. This shows one of the main advantages of the BAS: It permits the comparison of theories of different complexities, and naturally weighs the theories by their quality ({\em i.e.}, stating that nothing is known about $\delta$ is a very bad theory, so any not too bad approximation should be considered an improvement).

\begin{figure}
\includegraphics[scale=.4]{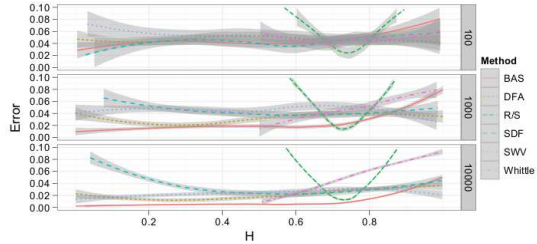}
\caption{\label{fig:allerror} (color online) Comparison of the quality of Hurst exponent estimators for 1,000 artificially generated fGn series of length 100 (top), 1,000 (middle), and 10,000 (bottom).}
\end{figure}

Fig.~\ref{fig:allerror} compares the accuracy of the estimator proposed here with some of the most commonly used estimators, including Rescaled Range Analysis (R/S; \cite{Hurst:1951,Mandelbrot:Wallis:1969}), Detrended Fluctuation Analysis ({DFA}; \cite{Peng:etal:1994}), the local Whittle estimator \cite{Robinson:1995}, and the Scaled Windowed Variance and Spectral Power Density ({SWV} and {SPD}; in their respective improved variants {SSC} and $^\mathrm{low}$SPD$_{w,e}$ as recommended by Ref.~\cite{Eke:etal:2002}). Already for series of mild length, the {BAS} estimator shows a substantially more accurate estimation than all other methods -- both in terms of error and consistency -- across all values $H < .9$. Only for $H > .9$ did the performace of the BAS estimator decrease, and even then it was not worse than the other estimators.

\begin{figure}
\includegraphics[scale=.35]{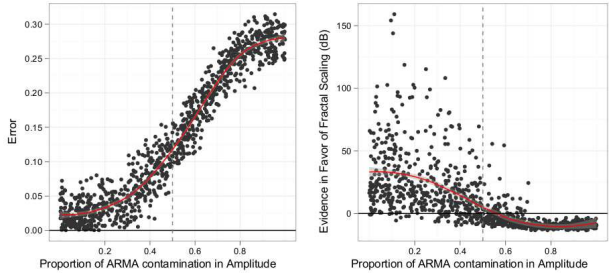}
\caption{\label{fig:arma} (color online) Robustness of the Hurst estimation (left) and BAS criterion (right) for 1,000 artificial fGn series ($\mathrm{H} = .75$) of length 1,000 with different proportions of contamination by an ARMA(2,1) process.}
\end{figure}

The next question that arises is that real world time series are hardly ever `clean' examples of FS, rather, the FS structure is usually corrupted by noise of different types. Of especial importance are short-term correlations, that some time can give the false impression of FS, or can mask it away. To investigate the robustness of the BAS to this type of signal contamination, I generated 1,000 fGn time series of length 1,000 with $\mathrm{H} = .75$ that were contaminated by random samples from a simulated $\mathrm{ARMA}(2,1)$ process (Autoregressive Moving Average; \cite{Hannan:Rissanen:1982}). The proportion of signal amplitude driven by ARMA was randomly sampled from $(0,1)$. Fig.~\ref{fig:arma} plots the resulting evidence (comparing $\delta\in[0,1]$ with $\delta=1/2$) and the error in estimating $\delta$ for each series. The {BAS} robustly detects the presence of FS up to the point when the ARMA component accounts for more than half of the signal power. The contamination drives down $\hat{\delta}$, and when this estimator goes below the $.61$ limit mentioned above, a plain short-term correlation provides a better account of the data.  

I have introduced an accurate method for deciding whether it is justified to claim that an observed series indicates FS.  The {BAS} is fully analytical, does not have parameters that need to be subjectively adjusted, and its results can be computed with at most one level of very constrained approximation (the Gaussian proposal).  The main component of the {BAS}, summarized by Eq.~\ref{eq:ll12}, is valid for stationary series of any type, whether they have finite variance, or they produce anomalous scaling of the L\'evy type. Additionally, if the values are roughly normally distributed (or can be normalized) one can use the normal approximation and associated H estimator. Combined, the estimator and the BAS provide a powerful tool, enabling the use of the estimator for generating hypotheses on one portion of the data, and the BAS criterion to test them on a different portion. If two theories make different scaling predictions, the BAS can be applied to decide which theory the data support. The BAS is only valid for stationary `noises'. If the series of interest is of the fractional Brownian motion type, differentiation is required before the methods can be applied. For more complex non-stationarity, preprocessing to isolate the stationary components of the series is required (see \cite{Ignaccolo:etal:2004a}, for a recent proposal). Finally, if it is not clear that the scaling should apply at the earliest times \cite{Ignaccolo:etal:2004a}, a later reference time $k_0 \geq 2$ can be chosen, and the time used in the equations expressed relative to it $k_r = k / k_0$.


\begin{thebibliography}{10}%
\makeatletter
\providecommand \@ifxundefined [1]{%
 \ifx #1\undefined \expandafter \@firstoftwo
 \else \expandafter \@secondoftwo
\fi
}%
\providecommand \@ifnum [1]{%
 \ifnum #1\expandafter \@firstoftwo
 \else \expandafter \@secondoftwo
\fi
}%
\providecommand \enquote [1]{``#1''}%
\providecommand \bibnamefont  [1]{#1}%
\providecommand \bibfnamefont [1]{#1}%
\providecommand \citenamefont [1]{#1}%
\providecommand\href[0]{\@sanitize\@href}%
\providecommand\@href[1]{\endgroup\@@startlink{#1}\endgroup\@@href}%
\providecommand\@@href[1]{#1\@@endlink}%
\providecommand \@sanitize [0]{\begingroup\catcode`\&12\catcode`\#12\relax}%
\@ifxundefined \pdfoutput {\@firstoftwo}{%
 \@ifnum{\z@=\pdfoutput}{\@firstoftwo}{\@secondoftwo}%
}{%
 \providecommand\@@startlink[1]{\leavevmode}%
 \providecommand\@@endlink[0]{}%
}{%
 \providecommand\@@startlink[1]{%
  \leavevmode
  \pdfstartlink
   attr{/Border[0 0 1 ]/H/I/C[0 1 1]}%
   user{/Subtype/Link/A<</Type/Action/S/URI/URI(#1)>>}%
  \relax
 }%
 \providecommand\@@endlink[0]{\pdfendlink}%
}%
\providecommand \url  [0]{\begingroup\@sanitize \@url }%
\providecommand \@url [1]{\endgroup\@href {#1}{\urlprefix}}%
\providecommand \urlprefix [0]{URL }%
\providecommand \Eprint[0]{\href }%
\@ifxundefined \urlstyle {%
  \providecommand \doi [1]{doi:\discretionary{}{}{}#1}%
}{%
  \providecommand \doi [0]{doi:\discretionary{}{}{}\begingroup
  \urlstyle{rm}\Url }%
}%
\providecommand \doibase [0]{http://dx.doi.org/}%
\providecommand \Doi[1]{\href{\doibase#1}}%
\providecommand \bibAnnote [3]{%
  \BibitemShut{#1}%
  \begin{quotation}\noindent
    \textsc{Key:}\ #2\\\textsc{Annotation:}\ #3%
  \end{quotation}%
}%
\providecommand \bibAnnoteFile [2]{%
  \IfFileExists{#2}{\bibAnnote {#1} {#2} {\input{#2}}}{}%
}%
\providecommand \typeout [0]{\immediate \write \m@ne }%
\providecommand \selectlanguage [0]{\@gobble}%
\providecommand \bibinfo [0]{\@secondoftwo}%
\providecommand \bibfield [0]{\@secondoftwo}%
\providecommand \translation [1]{[#1]}%
\providecommand \BibitemOpen[0]{}%
\providecommand \bibitemStop [0]{}%
\providecommand \bibitemNoStop [0]{.\EOS\space}%
\providecommand \EOS [0]{\spacefactor3000\relax}%
\providecommand \BibitemShut [1]{\csname bibitem#1\endcsname}%
\bibitem{Hurst:1951}%
  \BibitemOpen
  \bibfield{author}{%
  \bibinfo {author} {\bibfnamefont{H.~E.}\ \bibnamefont{Hurst}},\ }%
  \bibfield{journal}{%
  \bibinfo {journal} {Trans. Am. Soc. Civ. Eng.}\ }%
  \textbf{\bibinfo {volume} {116}},\ \bibinfo {pages} {770} (\bibinfo {year}
  {1951})%
  \bibAnnoteFile{NoStop}{Hurst:1951}%
\bibitem{Mandelbrot:VanNess:1968}%
  \BibitemOpen
  \bibfield{author}{%
  \bibinfo {author} {\bibfnamefont{B.~B.}\ \bibnamefont{Mandelbrot}}\ and\
  \bibinfo {author} {\bibfnamefont{J.~W.}\ \bibnamefont{{van~Ness}}},\ }%
  \bibfield{journal}{%
  \bibinfo {journal} {{SIAM} Rev.}\ }%
  \textbf{\bibinfo {volume} {10}},\ \bibinfo {pages} {422} (\bibinfo {year}
  {1968})%
  \bibAnnoteFile{NoStop}{Mandelbrot:VanNess:1968}%
\bibitem{Mandelbrot:Wallis:1969}%
  \BibitemOpen
  \bibfield{author}{%
  \bibinfo {author} {\bibfnamefont{B.~B.}\ \bibnamefont{Mandelbrot}}\ and\
  \bibinfo {author} {\bibfnamefont{J.~R.}\ \bibnamefont{Wallis}},\ }%
  \bibfield{journal}{%
  \bibinfo {journal} {Water Resources Research}\ }%
  \textbf{\bibinfo {volume} {5}},\ \bibinfo {pages} {228–} (\bibinfo {year}
  {1969})%
  \bibAnnoteFile{NoStop}{Mandelbrot:Wallis:1969}%
\bibitem{Eke:etal:2002}%
  \BibitemOpen
  \bibfield{author}{%
  \bibinfo {author} {\bibfnamefont{A.}~\bibnamefont{Eke}}, \bibinfo {author}
  {\bibfnamefont{P.}~\bibnamefont{Herm\'an}}, \bibinfo {author}
  {\bibfnamefont{L.}~\bibnamefont{Kocsis}},\ and\ \bibinfo {author}
  {\bibfnamefont{L.~R.}\ \bibnamefont{Kozak}},\ }%
  \bibfield{journal}{%
  \bibinfo {journal} {Physiol. Meas.}\ }%
  \textbf{\bibinfo {volume} {23}},\ \bibinfo {pages} {R1} (\bibinfo {year}
  {2002})%
  \bibAnnoteFile{NoStop}{Eke:etal:2002}%
\bibitem{Jaynes:2003}%
  \BibitemOpen
  \bibfield{author}{%
  \bibinfo {author} {\bibfnamefont{E.~T.}\ \bibnamefont{Jaynes}},\ }%
  \emph{\bibinfo {title} {Probability {T}heory: {T}he {L}ogic of {S}cience}}\
  (\bibinfo {publisher} {Cambridge {U}niversity {P}ress},\ \bibinfo {address}
  {Cambridge, UK},\ \bibinfo {year} {2003})%
  \bibAnnoteFile{NoStop}{Jaynes:2003}%
\bibitem{Davies:Harte:1987}%
  \BibitemOpen
  \bibfield{author}{%
  \bibinfo {author} {\bibfnamefont{R.~B.}\ \bibnamefont{Davies}}\ and\ \bibinfo
  {author} {\bibfnamefont{D.~S.}\ \bibnamefont{Harte}},\ }%
  \bibfield{journal}{%
  \Doi{10.1093/biomet/74.1.95}{\bibinfo {journal} {Biometrika}}\ }%
  \textbf{\bibinfo {volume} {74}},\ \bibinfo {pages} {95} (\bibinfo {year}
  {1987})%
  \bibAnnoteFile{NoStop}{Davies:Harte:1987}%
\bibitem{Wagenmakers:etal:2004}%
  \BibitemOpen
  \bibfield{author}{%
  \bibinfo {author} {\bibfnamefont{E.-J.}\ \bibnamefont{Wagenmakers}}, \bibinfo
  {author} {\bibfnamefont{S.}~\bibnamefont{Farrell}},\ and\ \bibinfo {author}
  {\bibfnamefont{R.}~\bibnamefont{Ratcliff}},\ }%
  \bibfield{journal}{%
  \bibinfo {journal} {Psychon. {B}ul. {R}ev.}\ }%
  \textbf{\bibinfo {volume} {11}},\ \bibinfo {pages} {579} (\bibinfo {year}
  {2004})%
  \bibAnnoteFile{NoStop}{Wagenmakers:etal:2004}%
\bibitem{Wagenmakers:etal:2005}%
  \BibitemOpen
  \bibfield{author}{%
  \bibinfo {author} {\bibfnamefont{E.-J.}\ \bibnamefont{Wagenmakers}}, \bibinfo
  {author} {\bibfnamefont{S.}~\bibnamefont{Farrell}},\ and\ \bibinfo {author}
  {\bibfnamefont{R.}~\bibnamefont{Ratcliff}},\ }%
  \bibfield{journal}{%
  \bibinfo {journal} {J. {E}xp. {P}sychol. {G}en.}\ }%
  \textbf{\bibinfo {volume} {134}},\ \bibinfo {pages} {108} (\bibinfo {year}
  {2005})%
  \bibAnnoteFile{NoStop}{Wagenmakers:etal:2005}%
\bibitem{Thornton:Gilden:2005}%
  \BibitemOpen
  \bibfield{author}{%
  \bibinfo {author} {\bibfnamefont{T.~L.}\ \bibnamefont{Thornton}}\ and\
  \bibinfo {author} {\bibfnamefont{D.~L.}\ \bibnamefont{Gilden}},\ }%
  \bibfield{journal}{%
  \bibinfo {journal} {Psychon. {B}ul. {R}ev.}\ }%
  \textbf{\bibinfo {volume} {12}},\ \bibinfo {pages} {409} (\bibinfo {year}
  {2005})%
  \bibAnnoteFile{NoStop}{Thornton:Gilden:2005}%
\bibitem{Peng:etal:1994}%
  \BibitemOpen
  \bibfield{author}{%
  \bibinfo {author} {\bibfnamefont{C.}~\bibnamefont{Peng}}, \bibinfo {author}
  {\bibfnamefont{S.~V.}\ \bibnamefont{Buldyrev}}, \bibinfo {author}
  {\bibfnamefont{S.}~\bibnamefont{Havlin}}, \bibinfo {author}
  {\bibfnamefont{M.}~\bibnamefont{Simons}}, \bibinfo {author}
  {\bibfnamefont{H.~E.}\ \bibnamefont{Stanley}},\ and\ \bibinfo {author}
  {\bibfnamefont{A.~L.}\ \bibnamefont{Goldberger}},\ }%
  \bibfield{journal}{%
  \Doi{{10.1103/PhysRevE.49.1685}}{\bibinfo {journal} {Phys. Rev. E}}\ }%
  \textbf{\bibinfo {volume} {49}},\ \bibinfo {pages} {1685} (\bibinfo {year}
  {1994})%
  \bibAnnoteFile{NoStop}{Peng:etal:1994}%
\bibitem{Allegrini:etal:1996}%
  \BibitemOpen
  \bibfield{author}{%
  \bibinfo {author} {\bibfnamefont{P.}~\bibnamefont{Allegrini}}, \bibinfo
  {author} {\bibfnamefont{P.}~\bibnamefont{Grigolini}},\ and\ \bibinfo {author}
  {\bibfnamefont{B.~J.}\ \bibnamefont{West}},\ }%
  \bibfield{journal}{%
  \Doi{10.1103/PhysRevE.54.4760}{\bibinfo {journal} {Phys. Rev. E}}\ }%
  \textbf{\bibinfo {volume} {54}},\ \bibinfo {pages} {4760} (\bibinfo {year}
  {1996})%
  \bibAnnoteFile{NoStop}{Allegrini:etal:1996}%
\bibitem{Scafetta:Grigolini:2002}%
  \BibitemOpen
  \bibfield{author}{%
  \bibinfo {author} {\bibfnamefont{N.}~\bibnamefont{Scafetta}}\ and\ \bibinfo
  {author} {\bibfnamefont{P.}~\bibnamefont{Grigolini}},\ }%
  \bibfield{journal}{%
  \Doi{{10.1103/PhysRevE.66.036130}}{\bibinfo {journal} {Phys. Rev. E}}\ }%
  \textbf{\bibinfo {volume} {66}},\ \bibinfo {pages} {036130} (\bibinfo {year}
  {2002})%
  \bibAnnoteFile{NoStop}{Scafetta:Grigolini:2002}%
\bibitem{Ignaccolo:etal:2004a}%
  \BibitemOpen
  \bibfield{author}{%
  \bibinfo {author} {\bibfnamefont{M.}~\bibnamefont{Ignaccolo}}, \bibinfo
  {author} {\bibfnamefont{P.}~\bibnamefont{Allegrini}}, \bibinfo {author}
  {\bibfnamefont{P.}~\bibnamefont{Grigolini}}, \bibinfo {author}
  {\bibfnamefont{P.}~\bibnamefont{Hamilton}},\ and\ \bibinfo {author}
  {\bibfnamefont{B.~J.}\ \bibnamefont{West}},\ }%
  \bibfield{journal}{%
  \Doi{10.1016/j.physa.2003.12.034}{\bibinfo {journal} {Physica A}}\ }%
  \textbf{\bibinfo {volume} {336}},\ \bibinfo {pages} {595} (\bibinfo {year}
  {2004})%
  \bibAnnoteFile{NoStop}{Ignaccolo:etal:2004a}%
\bibitem{Note1}%
  \BibitemOpen
  \bibinfo {note} {The case comparing two composite hypotheses can be derived
  trivially from Eq.~\ref {eq:ll12}.}%
  \bibAnnoteFile{Stop}{Note1}%
\bibitem{Kullback:Leibler:1951}%
  \BibitemOpen
  \bibfield{author}{%
  \bibinfo {author} {\bibfnamefont{S.}~\bibnamefont{Kullback}}\ and\ \bibinfo
  {author} {\bibfnamefont{R.~A.}\ \bibnamefont{Leibler}},\ }%
  \bibfield{journal}{%
  \bibinfo {journal} {Ann. Math. Stat.}\ }%
  \textbf{\bibinfo {volume} {22}},\ \bibinfo {pages} {79} (\bibinfo {year}
  {1951})%
  \bibAnnoteFile{NoStop}{Kullback:Leibler:1951}%
\bibitem{Note2}%
  \BibitemOpen
  \bibinfo {note} {Eq.~\ref {eq:step1} is transformed into Eq.~\ref {eq:step2}
  by a simple change of variable $z = y/k^{\delta },\protect \tmspace
  +\thinmuskip {.1667em} \protect \tmspace +\thinmuskip {.1667em}\protect
  \mathrm {d}y = k^{\delta } \protect \tmspace +\thinmuskip {.1667em}\protect
  \mathrm {d}z$.}%
  \bibAnnoteFile{Stop}{Note2}%
\bibitem{Robinson:1995}%
  \BibitemOpen
  \bibfield{author}{%
  \bibinfo {author} {\bibfnamefont{P.~M.}\ \bibnamefont{Robinson}},\ }%
  \bibfield{journal}{%
  \bibinfo {journal} {Ann. Stat.}\ }%
  \textbf{\bibinfo {volume} {23}},\ \bibinfo {pages} {1630} (\bibinfo {year}
  {1995})%
  \bibAnnoteFile{NoStop}{Robinson:1995}%
\bibitem{Hannan:Rissanen:1982}%
  \BibitemOpen
  \bibfield{author}{%
  \bibinfo {author} {\bibfnamefont{E.~J.}\ \bibnamefont{Hannan}}\ and\ \bibinfo
  {author} {\bibfnamefont{J.}~\bibnamefont{Rissanen}},\ }%
  \bibfield{journal}{%
  \bibinfo {journal} {Biometrika}\ }%
  \textbf{\bibinfo {volume} {69}},\ \bibinfo {pages} {81} (\bibinfo {year}
  {1982})%
  \bibAnnoteFile{NoStop}{Hannan:Rissanen:1982}%
\end{thebibliography}

%

\end{document}